\documentclass[12pt]{article}
\usepackage{latexsym}
\usepackage{hyperref}

\def\b{\beta}

\def\d{\delta}
\def\e{\epsilon}                
\def\g{\gamma}

\def\m{\mu}

\def\s{\sigma}                  
\def\t{\tau}

\def\D{\Delta}

\def\L{\Lambda}

\def\cf{{\cal F}}

\def\cl{{\cal L}}
\def\cm{{\cal M}}

\def\cbo{{\,\raise-.15ex\Sc [\,}}                       

\def\svev#1{\left\langle #1\right\rangle}       

\def\ddt#1{{\buildrel {\hbox{\LARGE .\kern-2pt.}} \over {#1}}}

\def\sstyle{\scriptstyle}

\def\ie{\mbox{\it i.e.} }
\def\eg{\mbox{\it e.g.} }

\def\frac#1#2{ {\sstyle {#1\over #2} } }

\thicklines                         
\def\str{{\rm str}}
\def\cf{{\it cf.}}

\begin{document}
\rightline{KANAZAWA-05-05}
\rightline{RBRC-489}
\begin{center}
\vspace{10mm}
{\large\bf The role of the double pole in lattice QCD\\ with mixed actions}
\\[12mm]
Maarten Golterman,$^a$\ \ Taku Izubuchi$^b$\ \ and \ \ Yigal Shamir$^c$
\\[8mm]
{\small\it
$^a$Department of Physics and Astronomy,
San Francisco State University\\
San Francisco, CA 94132, USA}\\
{\tt maarten@stars.sfsu.edu}
\\[5mm]
{\small\it $^b$RIKEN-BNL Research Center, Brookhaven National
Laboratory\\
Upton, NY 11973, USA, and\\
Institute for Theoretical Physics, Kanazawa University\\
Kakuma, Kanazawa, 920-1192, Japan}
\\
{\tt izubuchi@quark.phy.bnl.gov}
\\[5mm]
{\small\it $^c$School of Physics and Astronomy\\
Raymond and Beverly Sackler Faculty of Exact Sciences\\
Tel-Aviv University, Ramat~Aviv,~69978~ISRAEL}\\
{\tt shamir@post.tau.ac.il}
\\[10mm]
{ABSTRACT}
\\[2mm]
\end{center}

\begin{quotation}
We investigate effects resulting from the use of different discretizations
for the valence and the sea quarks in lattice QCD, considering
Wilson and/or Ginsparg--Wilson fermions.
We assume that such effects appear through scaling
violations that can be studied using effective lagrangian techniques.
We show that a double pole is
present in flavor-neutral Goldstone meson propagators,
even if the charged Goldstone mesons made out
of valence quarks and those made out of sea quarks have equal masses.
We then consider some observables known to be anomalously sensitive to
the presence of a double pole. For these observables, we find that the
double-pole enhanced scaling violations may turn out to be
rather small in practice.
\end{quotation}

\newpage
\section{Introduction}

Advanced numerical computations in Lattice QCD have recently seen an
increase in the use of mixed actions.  With ``mixed action" we refer
to any lattice theory in which differently discretized fermions are used in the
valence- and sea-quark sectors.  This development originates
from the desire to use a relatively affordable method to simulate the sea quarks,
using for instance various kinds of Wilson-like or staggered fermions, combined with a
need for very good chiral symmetry in the valence sector, which can be
obtained by using staggered or Ginsparg--Wilson quarks.

The worry about mixed actions is that they do not satisfy some
of the basic properties of a healthy quantum-field theory, such as
the possibility to analytically continue to Minkowski space and unitarity.
We recall that the same is true for quenched and partially quenched (PQ)
theories.  However, the PQ theory reduces to full QCD if valence- and sea-quark
masses are taken equal.  In the mixed-action case, no such limit exists
at non-zero lattice spacing.

In this paper we will make the following working assumptions.
(1) Any new unphysical effects resulting
from the use of a mixed action disappear in the continuum limit;
(2) At finite lattice spacing, these effects
are controlled by positive powers of the lattice spacing~$a$,
\ie they only show up through systematically manageable scaling violations;
(3) One can use effective-field theory techniques to investigate their size.

Effective field theory has proven to be a very useful tool for investigating
the unphysical effects resulting from quenching
and partial quenching, and
we expect the same for any new effects resulting from the use of mixed actions.
The idea is then simply to look for observables which are expected to be
maximally affected by the scaling violations of a mixed action.
Our experience with PQ theories gives us a clue for where to look.
When the valence-quark masses are chosen unequal to the sea-quark masses,
the most infrared-sensitive unphysical effects
originate in the double pole that occurs in flavor-neutral Goldstone-meson
two-point functions \cite{bg94}.

More generally,
the double pole is an unphysical effect which arises from choosing the
valence-quark action unequal to the sea-quark action.
Other unphysical effects are known to arise from
lattice-fermion actions containing beyond nearest-neighbor couplings,
and from gauge actions containing Wilson loops larger than a single plaquette.
These actions often lack reflection positivity,
and, hence, unitarity at finite lattice spacing.
Under such circumstances it is imperative to take the continuum limit
before analytically continuing to Minkowski space.
It is generally believed that the notion of universality applies to
the continuum limit obtained this way: one expects the same continuum limit
regardless of the details of the regularization, \ie of the
lattice discretization.   We expect that universality extends to mixed actions,
and our working assumptions translate this expectation
into more concrete terms. A practical test then lies in
comparing the results we obtain using chiral perturbation theory to numerical
lattice calculations. Agreement would indicate
that the assumptions underlying our effective-field theory calculations
are valid.

We will consider three classes of mixed actions, namely those with
Wilson-like (Wilson, clover, twisted-mass Wilson) sea
and valence quarks, those with Wilson-like sea quarks and Ginsparg--Wilson (GW)
valence quarks, and those with GW valence and sea quarks.
The latter refers to any lattice fermion satisfying
the GW relation~\cite{GW} either exactly or to a good approximation; overlap,
domain-wall, and perfect-action fermions included.

In all cases, we will
assume that different discretizations have been chosen in the
valence and sea sectors, \ie that the lattice Dirac operators for the valence-
and the sea-quarks differ not only in their bare mass parameters
(this includes for instance smearing the links
in the valence sector but not in the sea sector).
For simplicity we will restrict the discussion to degenerate masses
in the sea and valence sectors. Our results can be generalized
straightforwardly to non-degenerate and twisted quark-mass matrices.

The chiral lagrangian for mixed-action lattice QCD with Wilson-like and GW fermions
has been considered before in
Refs.~\cite{baeretal,baeretal2}, which we take as our starting point.\footnote{For
very recent work considering GW valence quarks with a staggered sea, see
Ref.~\cite{bbrs}.  Here we will not consider staggered fermions.}
In Sec.~2 we discuss the relevant elements of the chiral effective
theory. We calculate the residue of the double pole in the two-point
function of flavor-neutral Goldstone mesons, including the leading-order
scaling violations.  We then derive our main result:
In a mixed-action theory there is a double pole in this two-point function,
even if the charged Goldstone mesons made out of valence quarks and those
made out of
sea quarks have equal masses. If the sea and/or the valence action
is Wilson-like, the residue of the double pole is of order $a^2$.

We then go on in Sec.~3 to consider the size of this effect for some interesting
physical quantities which are particularly sensitive to the anomalous
infrared behavior originating from a double pole.
In particular, we consider the $I=0$ two-pion
energy shift in finite volume, where the double pole causes an anomalous
enhancement as a function of the volume \cite{bg96}, as well as the two-point
function of the $I=1$ scalar meson (the $a_0$), which has anomalous behavior
for large euclidean time \cite{bardeenetal}.\footnote{Similar
effects also occur in $K^0\to\pi^+\pi^-$ matrix elements \cite{gp}.}  Our conclusions
are in Sec.~4.

\section{The residue of the double pole in mixed theories}

Our analysis starts from the results of Refs.~\cite{baeretal,baeretal2}.
We first consider the implementation of both valence and sea quarks
through Wilson-like fermions.
``Wilson-like" stands for any lattice fermion with exact flavor symmetry,%
\footnote{As already mentioned, for simplicity we limit the discussion
  to degenerate masses in the sea and valence sectors.
  The generalization to non-degenerate \cite{bg94} and twisted \cite{tmpq}
  masses is straightforward.}
but with no chiral symmetry at non-zero lattice spacing $a$.
In Ref.~\cite{baeretal2} first the Symanzik
effective quark action was constructed\footnote{For a review of the method, as well as further
references, see Ref.~\cite{baer}.} for Wilson fermions to order $a^2$.  If both valence
and sea quarks are made of the same type of Wilson-like fermion,
the (vector) symmetry group
is that of a partially-quenched (PQ) theory \cite{bg94}, $SU(K+N|K)$, where $K$ is the
number of valence quarks, and $N$ is the number of sea quarks.  In the continuum limit
chiral symmetry is restored
(for massless quarks), and the symmetry group enlarges to\footnote{For
explanations about the legitimacy of using this symmetry group, see Refs.~\cite{ShSi,GSS}.}
\begin{equation}
SU(K+N|K)_L\times SU(K+N|K)_R\,.\ \ \ \ {\rm (partially\ quenched)} \label{pq}
\end{equation}

Ref.~\cite{baeretal2} also considered Wilson-like
sea quarks with valence quarks of the GW type \cite{GW,MLsymm}.
The $a\ne 0$ mixed-action theory is then invariant under
\begin{equation}
[SU(K|K)_L\times SU(K|K)_R]\times SU(N)\,. \ \ \ {\rm (GW-Wilson)} \label{GWW}
\end{equation}
Again the complete Symanzik effective action was constructed to order $a^2$.
For both cases,
the complete chiral lagrangian was derived to the same order from the
Symanzik effective action with the help of a spurion analysis.

Here we are only interested in the role of the double pole in neutral Goldstone meson
propagators, and we
will restrict ourselves to operators in the chiral lagrangian relevant for our purpose.
However, our aim is to consider also mixed actions
in which both valence and sea quarks
are Wilson-like, but different.
In that case, the symmetry group is reduced to
\begin{equation}
SU(K|K)\times SU(N)\,, \ \ \ {\rm (Wilson-Wilson)} \label{WW}
\end{equation}
and more operators  appear in the effective theory.

In the continuum limit, the chiral effective lagrangian is just that of PQ QCD \cite{bg94}.
In terms of the non-linear field $\Sigma={\rm exp}(2i\Phi/f)$, with $\Phi$ the matrix of
Goldstone-meson fields, the lowest-order (euclidean) chiral lagrangian is
\begin{equation}
\cl_{\rm cont}={1\over 8}\, f^2\,\str(\partial_\m\Sigma\partial_\m\Sigma^\dagger)-
{1\over 4}\,f^2 B_0\,\str(\Sigma \cm^\dagger+\cm\Sigma^\dagger)+{1\over 2}\,\mu^2\,\Phi_0^2\,,\label{chlag}
\end{equation}
where $\Phi_0=\str(\Phi)=(f/2i)\log{{\rm sdet}(\Sigma)}$, and $f$, $B_0$ and $\mu^2$ are low-energy constants (LECs).\footnote{Note that $\mu^2({\rm this\ paper}) =\mu^2({\rm
Ref.~\cite{bg94}})/3$.}
$\cm$ is the quark mass matrix,
\begin{equation}
\cm=\cm^\dagger={\rm diag}(m_v,m_v,\dots,m_s,m_s,\dots,m_v,m_v,\dots)
\,,\label{mass}
\end{equation}
while $\str$ and ${\rm sdet}$ are the supertrace and the superdeterminant
needed to construct invariants for graded groups.
The first $K$ masses are those of the valence quarks, the next $N$ those of the sea
quarks, and the final $K$ those of the ghost quarks needed to cancel the loops of the
valence quarks \cite{morel}.

In Eq.~(\ref{chlag}) we kept the singlet mass term because we are interested in
scaling violations to the singlet mass parameter $\m^2$. Normally one would not
do this because $\mu$ is considered as being of same order as
the chiral-lagrangian cutoff scale $\L\approx 4\pi f$.
The pseudo-scalar singlet's mass-squared
is equal to $N\mu^2+2B_0 m_s$, and so
this degree of freedom can be integrated out.  Accordingly, one could constrain
$\Sigma$ by setting ${\rm sdet}(\Sigma)=1$, \ie $\Phi_0=0$.
We will indeed recover this result in what follows,
and from then on limit ourselves to the theory with the pseudo-scalar singlet integrated
out.  This is also the reason that we did not keep any other terms containing the singlet
field $\Phi_0$ in Eq.~(\ref{chlag}).

Because of the reduced symmetry group~(\ref{WW}) for a Wilson-like mixed-action,
the following operators appear at order $a^2$ in the chiral
lagrangian:
\begin{eqnarray}
\d\cl_{\rm W}&=&-{(af)^2\over 32}\left(\g_{vv}\,(\str(P_v(\Sigma-\Sigma^\dagger)))^2
+\g_{ss}\,(\str(P_s(\Sigma-\Sigma^\dagger)))^2\right.\nonumber\\
&&\hspace{1.7cm}\left.+2\g_{vs}\,\str(P_v(\Sigma-\Sigma^\dagger))\,\str(P_s(\Sigma-\Sigma^\dagger))\right)\,.
\label{LWW}
\end{eqnarray}
Here $P_v$ and $P_s$ are projectors on the valence, respectively, sea sectors,
\begin{equation}
P_v={\rm diag}({\bf 1}_v, 0, {\bf 1}_g)\,,\ \ \ \ \
P_s={\rm diag}(0,{\bf 1}_s, 0)\,, \label{proj}
\end{equation}
where ${\bf 1}_v$ and ${\bf 1}_g$ are the $K\times K$ identity matrices
in respectively the valence and ghost sectors, and
${\bf 1}_s$ the $N\times N$ identity matrix in the sea sector.
Eq.~(\ref{LWW}) follows from an
analysis very similar to that of Ref.~\cite{baeretal2}.
It was shown there that the corresponding operators
in the Symanzik effective action are four-fermion
operators, which appear at order $a^2$.
As usual, ``translation'' to the chiral effective lagrangian is done
using a spurion analysis. Here the spurions are needed to restore
both chiral symmetry and full valence--sea symmetry.
When expectation values are assigned to them, both these symmetries
are explicitly broken, {\it cf.} Eq.~(\ref{LWW}).

In Ref.~\cite{baeretal2} only one similar operator
appears in the chiral lagrangian,
because the same Wilson-fermion action was assumed for both the valence and sea quarks.
The symmetry group is then $SU(K+N|K)$ instead of (\ref{WW}),
which implies\footnote{This corresponds to the operator
with LEC $W'_7$ in Eq.~(21) of Ref.~\cite{baeretal2}.  The relevant spurions
are $B$ and $C$ of Eq.~(14) of Ref.~\cite{baeretal2}.
Replacing the identity matrix $I$ in that equation by $P_v$ or $P_s$
in all possible combinations leads to our Eq.~(\ref{LWW}).}
\begin{equation}
\g_{vv}=\g_{vs}=\g_{ss}\,.\ \ \ {\rm (partially\ quenched,\ Wilson)} \label{gammas}
\end{equation}

We will also consider mixed actions
in which the valence quarks are of the GW type,
while the sea quarks are Wilson-like \cite{baeretal2}.
The exact chiral symmetry in the valence sector (\cf\ Eq.~(\ref{GWW}))
implies $\g_{vv}=\g_{vs}=0$ in Eq.~(\ref{LWW}).
What survives is only the term corresponding entirely to the sea sector,%
\footnote{The corresponding operator is the one multiplying $W'_7$ in Eq.~(40)
of Ref.~\cite{baeretal2}.}
where chiral symmetry is still broken at $a\ne 0$.

Finally, when both valence and sea quarks are of the GW type, no such
operators can appear at all, and all three $\g$'s vanish, consistently
with the enlarged symmetry group (for massless quarks)
\begin{equation}
[SU(K|K)_L\times SU(K|K)_R]\times [SU(N)_L\times SU(N)_R]\,.\ \ \ \ {\rm (GW - GW)} \label{GWGW}
\end{equation}
Operators similar to those in Eq.~(\ref{LWW}) can appear only at order $a^2m^2$,
with $m\sim m_v\sim m_s$.
At the level of the chiral lagrangian, they can be made
by replacing the factors
$\str(P_{v,s}(\Sigma-\Sigma^\dagger))$ in Eq.~(\ref{LWW}) by
$\str(P_{v,s}(\Sigma\cm^\dagger-\cm\Sigma^\dagger))$,
where $\cm$ is now promoted to
a spurion field transforming like $\Sigma$.

For domain-wall fermions with a finite fifth dimension $L_5$,
chiral symmetry is only approximate \cite{fs}.
(A mixed action could result \eg from choosing different values for $L_5$
in the valence and sea sectors.)
The ensuing chiral-symmetry violating effects
can be taken into account by adding separate spurions for the residual
quark masses $m_{\rm res}^{v,s}$ \cite{rbc,gsmres}.
As long as $m_{\rm res}$ is smaller than the explicit quark masses,\footnote{
  If $L_5$ is chosen so small (\eg $L_5=2$) that the residual masses, and,
  hence, the actual quark masses are substantially bigger than the
  explicit quark-mass parameters,
  domain-wall fermions  should be treated as Wilson-like.
}
again all such operators are of order $a^2m^2$.

In order to obtain tree-level meson propagators,  we expand $\cl_{\rm cont}+\d\cl_{\rm W}$ to quadratic order in the meson field $\Phi$.  It is
useful to define separate singlet fields in the valence and sea sectors:
the ``super-$\eta'$" $\Phi_0^v$ of the valence-symmetry group $SU(K|K)$ \cite{bg92},
and  the ``dynamical" singlet field $\eta'_s$. Explicitly,
\begin{equation}
\Phi_0^v\equiv\sum_{i\ {\rm valence}}\Phi_{ii}-\sum_{i\ {\rm ghost}}\Phi_{ii}\,,\ \ \ \ \
\eta'_s\equiv\sum_{i\ {\rm sea}}\Phi_{ii}\,,
\label{etadef}
\end{equation}
where $\Phi_0=\Phi_0^v+\eta'_s$.
The singlet mass terms following from Eqs.~(\ref{chlag},\ref{LWW}) can now be written as
\begin{equation}
\cl_{\rm singlet}=
{1\over 2}\,\mu_{vv}^2\,(\Phi_0^v)^2+\mu_{vs}^2\,\Phi_0^v\,\eta'_s+{1\over 2}\,\mu_{ss}^2\,(\eta'_s)^2\,,\label{etamass}
\end{equation}
in which
\begin{eqnarray}
\mu_{vv}^2&=&\mu^2+\g_{vv} a^2\,,\label{mus}\\
\mu_{vs}^2&=&\mu^2+\g_{vs} a^2\,,\nonumber\\
\mu_{ss}^2&=&\mu^2+\g_{ss} a^2\,.\nonumber
\end{eqnarray}
These scaling violations contain no derivatives, nor
(for Wilson fermions) powers of the quark mass matrix.
Hence they are part of the leading-order difference in the chiral expansion
between a mixed action at non-zero lattice
spacing and its PQ continuum limit.

Before proceeding, we have a few comments.  First, new operators as in Eq.~(\ref{LWW})
lead not only to order $a^2$ contributions to singlet mass terms, but also to new $O(a^2)$ vertices
in the chiral lagrangian.  In other words, chiral symmetry implies that Eq.~(\ref{etamass}) cannot appear as an isolated invariant term.\footnote{We thank Claude Bernard for very useful discussions on this
point.}  Second, while other operators appear at order $a^2$ in the chiral lagrangian \cite{baeretal2}, all the quadratic terms are of the form
\begin{equation}
a^2\left(\b_v\,\str(P_v\Phi^2)+\b_s\,\str(P_s\Phi^2)\right)\,,\label{othermass}
\end{equation}
with $\b_{v,s}$ linear combinations of order $a^2$ LECs.\footnote{In the GW-valence,
Wilson-sea case, $\b_s$ is a linear combination of $W'_6$ and $W'_8$ in Eq.~(40) of
Ref.~\cite{baeretal2}.}
They do not contribute any singlet mass terms, \ie terms as in Eq.~(\ref{etamass}).

With our quark mass matrix (\ref{mass}),
all pions made out of valence quarks are degenerate with a common mass
$M_{vv}$, and, similarly, pions made out of sea quarks all have a common mass $M_{ss}$
with, in general, $M_{ss}\ne M_{vv}$.   Here $M_{vv}$ denotes the mass of a charged
valence meson, \ie a meson made out of a quark and an anti-quark of different flavors.
At tree level, these meson masses are given by \cite{baeretal2}%
\footnote{We assume that $1/a$ divergences
have been absorbed into $m_{v,s}$.  We also assume that we are close to, but outside
the Aoki phase \cite{aokiphase}.}
\begin{eqnarray}
M_{vv}^2&=&2B_{0v}m_v+2W_{0v}a+2\b_v a^2+O(am,m^2,a^3)\,,\label{mesonmass}\\
M_{ss}^2&=&2B_{0s}m_s+2W_{0s}a+2\b_s a^2+O(am,m^2,a^3)\,,\nonumber
\end{eqnarray}
where $B_{0v}-B_{0s}=O(a^2)$.
The LECs $W_{0v}$ or $W_{0s}$ vanish if one works with an order-$a$ improved action
in the valence or sea sector.
The terms of order $a$ and $a^2$ both vanish if one uses fermions of the GW type.  A key observation
is that, at tree level, the operators in $\d\cl_{\rm W}$ only contribute to singlet mass terms, as in
Eq.~(\ref{etamass}). The LECs $\g_{vv,vs,ss}$ do not appear in the tree-level expressions
for the meson masses $M_{vv,ss}$.

We obtain the tree-level two-point function
$G_{i,j}=\svev{\Phi_{ii}\,\Phi_{jj}}$ for the neutral valence
meson fields as
\begin{equation}
G_{i,j}(p)={\d_{ij}\over p^2+M_{vv}^2}
-{{\mu_{vv}^2(p^2+M_{ss}^2)+N(\mu_{vv}^2\mu_{ss}^2-\mu_{vs}^4)}\over
{(p^2+M_{vv}^2)^2(p^2+M_{ss}^2+N\mu_{ss}^2)}}\,,\label{singletprop}
\end{equation}
where the indices $i$ and $j$ correspond to the flavor of the valence quark.
Since the effective theory is based on a combined
expansion in $M_{vv,ss}^2\sim p^2$ and $a^2$,
treating $\mu^2$ as a quantity of order one
in this expansion, we can simplify this expression to leading order in these
parameters by using Eq.~(\ref{mus}).
We obtain\footnote{Note that we do not specify the {\it relative} size of $M_{vv,ss}^2$ and $a^2$.}
 \begin{equation}
G_{i,j}(p)=\left(\d_{ij}-{1\over{N}}\right){1\over p^2+M_{vv}^2}
-\,{{(M_{ss}^2-M_{vv}^2)/N+(\g_{vv}+\g_{ss}-2\g_{vs})a^2}\over
{(p^2+M_{vv}^2)^2}}\,,\label{singletpropsimp}
\end{equation}
Indeed, $\mu^2$ disappears from the propagator, consistent with the fact that
in a PQ theory $\Phi_0=\str(\Phi)$ can be constrained to vanish
\cite{shsh}.  Setting $\Phi_0=0$ implies
that $\eta'_s=-\Phi_0^v$, and thus $\cl_{\rm singlet}$ goes over into
\begin{equation}
\cl_{\rm singlet}\to{1\over 2}\,a^2(\g_{vv}+\g_{ss}-2\g_{vs})(\Phi_0^v)^2\,.
\label{Lsingletsimp}
\end{equation}
This simplified singlet mass term directly leads to Eq.~(\ref{singletpropsimp}).
Setting $\str(\Phi)=0$ also removes terms of the form $\Phi_0\,\str(P_{v,s}
(\Sigma-\Sigma^\dagger))$, $(\partial_\mu\Phi_0)^2$, {\it etc.},
so that we were justified in not considering such terms in
Eq.~(\ref{LWW}).

{}From Eq.~(\ref{singletpropsimp}), we learn that in a mixed-action theory, double poles may arise
from two different sources.  One is well known from PQ QCD. It has a residue
proportional to the difference between valence- and sea-meson's mass-squared.
In addition, at non-zero lattice spacing, a new term proportional to
$a^2$ appears in the residue.
This new term arises because scaling violations in the valence, sea and
mixed double hairpins are different for such theories, and thus $\g_{vv}+\g_{ss}-2\g_{vs}$ will
in general not vanish.  Of course, if we use the same lattice fermions for both valence
and sea sectors, we recover the PQ case in which this new contribution to the residue
does vanish, {\it cf.} Eq.~(\ref{gammas}).

To the order we work, it is possible to tune the residue of the double pole
to zero by adjusting the valence quark mass.
A vanishing residue would be obtained by choosing
the valence quark mass $m_v$ such that the residue,
\begin{equation}
R=(M_{ss}^2-M_{vv}^2)/N+a^2(\g_{vv}+\g_{ss}-2\g_{vs})\,,\label{residue}
\end{equation}
vanishes. Alternatively, one could tune the valence quark mass
such that $M_{vv}=M_{ss}$. At non-zero lattice spacing,
these two choices are {\it different}.
In other words, while also the meson masses are
afflicted by scaling violations (\cf\ Eq.~(\ref{mesonmass})),
there are scaling violations in the double pole which do not get
absorbed into the meson masses.  If one chooses to tune the valence
quark mass such as to set $R=0$, obviously this requires a quantity sensitive
to the singlet part of the $\eta'$ mass-squared, $N\mu^2$.

Our analysis does not apply to any mixed action involving staggered
fermions.  Mixed actions with only staggered or GW-type
fermions have no order-$a$ scaling violations. Order-$a^2$ scaling
violations associated with symmetry breaking can only come from the breaking
of taste symmetry in the staggered-fermion sector.
As a result, all the order-$a^2$
scaling violations in the double pole can be related to the order-$a^2$
mass splittings between the staggered pseudo-scalar mesons \cite{bbrs}.
For mixed actions with staggered and/or GW-type fermions,
scaling violations that cannot be absorbed
into the Goldstone-meson masses occur in the residue of the double pole
only at order $a^2 m^2$.

\section{Examples of quantities with enhanced scaling violations}

The occurrence of a double pole in neutral-meson propagators
leads to an enhanced sensitivity of many quantities to the meson masses
\cite{bg92,sharpe92,bardeenetal,prelovseketal} or to the volume \cite{bg96,gp,damgaard}.  This
implies {\it ipso facto} an enhanced sensitivity to scaling violations of such
quantities because of the presence of the LECs $\g_{vv,ss,vs}$.  In this section we
discuss examples of such quantities, in particular the $I=0$ pion
scattering length \cite{bg96} and the two-point function for the $I=1$ scalar meson
$a_0$ \cite{bardeenetal,prelovseketal}.

The oldest examples of enhanced sensitivity due to the double pole are the
enhanced (``quenched") chiral logarithms in Goldstone meson masses, decay constants,
and the chiral condensate \cite{bg92,sharpe92}.
However, for quark masses used in typical simulations, these
effects are numerically very small; the same is expected to be true for the
scaling violations present in Eq.~(\ref{singletpropsimp}).  A more interesting quantity
is the shift in the energy of two pions, both at rest in an $I=0$ state,
in a finite spatial volume $L^3$ with periodic boundary conditions.
In quantities like this, the presence of a double pole gives rise to enhancement of
finite-volume effects \cite{bg96}.  It is therefore
{\it a priori} not clear what the combined effect is of enhancement by powers
of $L$ in addition to suppression by powers of $a$.

The two-pion $I=0$ energy
shift is given by \cite{bg96}
\begin{equation}
{\D E_{I=0}\over 2M_{vv}}=-{7\pi \over{8f^2M_{vv}L^3}}+{1\over 2}B_0(M_{vv} L)\d^2
+{1\over 2}\left(1-{1\over N}\right)
A_0(M_{vv} L)\d\e+O(\e^2)\,,\label{enshift}
\end{equation}
in which
\begin{equation}
\d\equiv{R\over 8\pi^2 f^2}\,,\ \ \ \ \ \e\equiv{M_{vv}^2\over 16\pi^2 f^2}\,,\label{abbr}
\end{equation}
and $R$ is defined in Eq.~(\ref{residue}).
The first term is the
leading-order, tree-level result, while the other terms come from one-loop diagrams.
We adapted the result of Ref.~\cite{bg96} to the PQ case, resulting in the
extra $-1/N$ appearing in the $A_0$ term.%
\footnote{In the fully quenched case, which corresponds to taking
$M_{ss}^2\to\infty$ in Eq.~(\ref{singletprop}), the corresponding factor
is equal to one.}

The large-$L$ expansions of the
coefficient functions $B_0$ and $A_0$ are given by\footnote{This corrects
a mistake in $B_0(ML)$ in Eq.~(18) of Ref.~\cite{bg96}.}
\begin{eqnarray}
B_0(ML)&=&-{5z(3)\over 8\pi^2}+{5z(2)\over 32\pi^2}\left({2\pi\over ML}\right)^2+
O\left(\left({2\pi\over ML}\right)^3\right)\,,
\label{BA}\\
A_0(ML)&=&{3z(2)\over 4\pi^2}\left({2\pi\over ML}\right)^2
+O\left(\left({2\pi\over ML}\right)^3\right)\,,
\nonumber
\end{eqnarray}
in which $z(3)$ and $z(2)$ are geometrical constants,
\begin{equation}
z(3)=8.40192397\,,\ \ \ \ \ z(2)=16.53231596\,.\label{zs}
\end{equation}
We only showed the terms that generate enhanced finite-volume corrections
to the leading-order result.  These
corrections are unphysical, and are absent if exactly the same
lattice-fermion action is used in the valence and sea sectors  (with valence-quark
masses equal to sea-quark masses).
When a mixed action is used, if one
does not tune $m_v$ to make the residue in Eq.~(\ref{residue}) vanish,
$\D E_{I=0}$ would be afflicted by scaling violations.
(As already explained,
this is true even if $m_v$ is tuned to set $M_{vv}=M_{ss}$.)
The enhanced scaling violations are
of order $M_{vv}a^4L^3\L_{\rm QCD}^6$ (from the leading term in $B_0$),
respectively $M_{vv}a^2L\L_{\rm QCD}^2$ (from the leading term in $A_0$),
relative to the physical value (estimating $f\sim \L_{\rm QCD}$ and
$\g_{vv,vs,ss}\sim \L_{\rm QCD}^4$).
There are additional four-point vertices of order $a^2$ which
contribute to these one-loop terms in $\D E_{I=0}$, that were not taken
into account in Ref.~\cite{bg96}.  However, they will not lead to any
parametrically larger contributions than the ones considered here.

It is instructive to consider these results in view of various power-counting schemes
proposed in the literature.
Ref.~\cite{baeretal2} considered an expansion
with a single
small parameter (compare Eq.~(\ref{abbr}))
$\e\sim M_{vv,ss}^2/\L^2\sim p^2/\L^2\sim a\L_{\rm QCD}$.
Assuming $M_{vv}L\gg 1$ and $M_{vv}=M_{ss}$, the first term of
Eq.~(\ref{enshift}) is of order $\e/(M_{vv}L)^3$, the second term is of order $\e^4$, while the third term
is of order $\e^3/(M_{vv}L)^2$.
If we would ignore the enhancement of the second and third terms by positive powers
of $M_{vv}L$ relative to the first term, these terms would be of
higher order compared to the $O(\e^2)$ term in Eq.~(\ref{enshift}) and
the $O(\e^2)$ terms considered in Ref.~\cite{baeretal2}.
(This is consistent with the fact that, with this power counting,
the $a^2$ term in $R$ is of higher order than the $M_{ss}^2-M_{vv}^2$ term.)
In a different
scheme \cite{aoki}, $(a\L_{\rm QCD})^2$ is taken to be of order $\e$.  In that case, the second and third terms in Eq.~(\ref{enshift}) are both of order $\e^2$.
Again they are
enhanced (by $(M_{vv}L)^3$, respectively $M_{vv}L$) relative to the other $O(\e^2)$ terms
in Eq.~(\ref{enshift}).

For parameters used in realistic simulations,
the volume-enhanced unphysical effects may nevertheless turn
out to be rather small.
For instance, setting $aM_{vv}\sim 0.2$, $a\L_{\rm QCD}\sim 0.1$
and $L/a=32$ leads to scaling violations of about 0.7\%
(from $B_0(M_{vv}L)$) and 6\% (from $A_0(M_{vv}L)$), relative to the first,
physical term in Eq.~(\ref{enshift}).  Note that the numerical
coefficient of $(M_{vv}L)^{-2}$ in $A_0(M_{vv}L)$ is rather large, about 50.

We see that, if Wilson-like fermions are used in the valence and/or
sea sectors, the unphysical effects in $\Delta E_{I=0}$ are expected to be rather small
in realistic simulations,
although probably not entirely negligible.
Of course, better knowledge of
$\g_{vv}+\g_{ss}-2\g_{vs}$ is needed in order to move beyond our crude
estimates.

As explained earlier, when GW fermions with exact
chiral symmetry are employed in both the valence and sea sectors,
the scaling violations (in $R$) are further suppressed by
a factor of order $(m/\L_{\rm QCD})^2$.
Thus, the unphysical effects will be negligibly small
if the quark masses are light enough.
The same is true for domain-wall fermions, provided the residual mass
is not larger than typical explicit quark masses
(as easily accomplished in current simulations).

Another quantity which may be strongly afflicted by the double
pole is the $I=1$ scalar ($a_0$) two-point function \cite{bardeenetal}.
This effect was considered in the PQ theory \cite{prelovseketal},
which we will generalize to the mixed-action case.
We note that the sensitivity of the $a_0$ two-point function
was used to tune a mixed action theory to the full unquenched QCD
limit in Ref.~\cite{ukqcd}.
However, the scaling violations present in the
double pole, Eq.~(\ref{singletpropsimp}), were not considered in that study.

The $I=1$ scalar two-point correlator at vanishing spatial momentum
was calculated in the PQ theory \cite{prelovseketal}.  Adapting that
result to our case, the dominant contribution for large
time $t$ is given by\footnote{Our LEC $B_0$ corresponds to $2\m_0$
of Ref.~\cite{prelovseketal}, and our $\m^2$ corresponds to $m_0^2/3$.}
\begin{equation}
C(t)\rightarrow {B_0^2\over 2L^3}\left[{e^{-2M_{vs}t}\over M_{vs}^2}\,{N\over 2}
-\,{e^{-2M_{vv}t}\over M_{vv}^2}{1\over M_{vv}^2}
\left((M_{vv}^2+M_{ss}^2)/N+RM_{vv}t\right)\right]\,.
\label{timecorr}
\end{equation}
Again, $R$ is the residue of the double pole
(no particular assumption on the value of $M_{ss}^2-M_{vv}^2$ is made).
$M_{vs}$ is the mass of a meson made out of a valence and a sea quark,
with $M_{vs}^2=(M_{vv}^2+M_{ss}^2)/2$ to leading order.  We have assumed that
the mass of the $\eta'_s$, given by $M_{\eta'}^2=N\m^2+M_{ss}^2$, as
well as the $a_0$ mass, are much heavier than any Goldstone-meson mass.
For large enough $t$ the linear term in $t$ dominates.
(Depending on the actual values of the $M$'s and $\g$'s in Eq.~(\ref{residue}),
$C(t)$ could even become negative.)
Setting
$M_{vv}=M_{ss}$ reduces Eq.~(\ref{timecorr}) to
\begin{equation}
C(t)\rightarrow {B_0^2\over 2L^3}\,{e^{-2M_{vv}t}\over M_{vv}^2}\,
\left({N\over 2}-{2\over N}-{(\g_{vv}+\g_{ss}-2\g_{vs})a^2t\over M_{vv}}
\right)\,.\label{domin}
\end{equation}
The coefficient of the factor linear in $t$ (measured in lattice units) is
of order $(a\L_{\rm QCD})^4/$ $(aM_{vv})$,
which is again quite small for a typical simulation.
(In the special case $N=2$ the first two terms in Eq.~(\ref{domin})
cancel each other, and the last term should be compared
with the $a_0$ and $\eta'$
terms which are present in the full correlator.)

\section{Conclusions}

We have investigated two quantities which are expected to be
particularly sensitive to the unphysical effects
resulting from a mixed lattice-QCD action.
We have assumed that there is no problem of principle, \ie
we have assumed that all such effects are encoded in scaling violations.
Still, it is important to estimate their numerical size for realistic choices
of MC simulation parameters. Using effective field theory
techniques, building on the work of Refs.~\cite{baeretal, baeretal2},
we identified the double pole in the flavor-neutral Goldstone meson
propagators as the most infrared-sensitive probe of these effects.

We considered two examples of observables where the double-pole
scaling violations are parametrically enhanced,
either by powers of $M_{vv}L$ (\ie of the spatial size)
in the case of $I=0$ two-pion energy shifts,
or linearly with euclidean time $t$ in the case of the $a_0$
propagator.
We found that these scaling violations are likely to
be rather small in currently realistic simulations.
While this is good news for anyone interested in using mixed actions,
care should be taken to monitor their size in practice.
Indeed, for the two-pion energy shift
the effect---while small---is not negligible already according to
our crude estimate, if a Wilson-like action is used in the sea and/or valence
sector.

The magnitude of these scaling violations depends on the observable
in question. It could well be that other interesting observables
exist for which the effect is larger.  As another example, let us
mention the nucleon-nucleon potential,
where the contribution from one-pion exchange is given by a Yukawa potential.
In PQ QCD with unequal valence- and sea-quark
masses, the long-distance part of the potential gets an additional
contribution from the double pole through $\eta$ exchange \cite{bs}.%
\footnote{We thank Paulo Bedaque for
reminding us of this example of quenching effects.}
Including the scaling violations of Eq.~(\ref{singletpropsimp}) leads
to a potential of the form
\begin{equation}
V(r)={1\over 8\pi f^2}\,\s_{(1)}\cdot\nabla\,\s_{(2)}\cdot\nabla\left(
g_A^2{\t_{(1)}\cdot\t_{(2)}\over r}-g_0^2{R\over M_{vv}}\right)\,e^{-M_{vv}r}\,,
\label{Yukawa}
\end{equation}
where we took $N=2$. Here $g_A$ ($g_0$) is the $I=1$ ($I=0$) axial coupling.
The indices $(1)$ and $(2)$ refer to the two nucleons, the $\s$'s to spin
and the $\t$'s to isospin.
The second, unphysical term clearly dominates the first term at large distances.
The nucleon-$\eta$ coupling $g_0$ is poorly known, but it could be
larger than $g_A$, thus decreasing the distance at which the unphysical term
sets in.

If lattice fermions satisfying the GW
relation (or approximations thereof such as domain-wall fermions with
finite fifth extent) are used in both the valence and sea sectors,
chiral symmetry leads to an extra suppression of the unphysical mixed-action
effects, by a factor of order $(m/\L_{\rm QCD})^2$.

We conclude with a practical comment.
For Goldstone-meson masses in the few 100~MeV range,
one has $M_{vv}/\L_{\rm QCD} \sim M_{ss}/\L_{\rm QCD} = O(1)$.
In comparison, $(a\L_{\rm QCD})^2 = O(0.01)$.
Therefore, when Wilson-like fermions are involved, one might have to control the
difference $M_{vv}^2 - M_{ss}^2$ at the few percent level
in order to be able to single out the scaling-violations part
of the residue $R$. This is likely to require the use of
an observable with enhanced sensitivity to the double pole.
If both sea and valence fermions are of the GW type
and the quark masses are small enough,
the effect could be too small to be detected.

\section*{Acknowledgements}

We thank Oliver B\"ar, Paulo Bedaque, Claude Bernard and Steve Sharpe
for helpful discussions.
This work was started during a visit of two of us (MG and YS) to
Brookhaven National Lab and Columbia University, and we thank the RBC collaboration
for discussions and hospitality.  MG also thanks the Lawrence Berkeley
National Lab Nuclear Theory Group for hospitality.
MG is supported in part by the US Department of Energy, and
YS is supported by the Israel Science Foundation under grant
222/02-1.

\end{document}